# Physics-Informed Machine Learning for EDFA: Parameter Identification and Gain Estimation

Xiaotian Jiang, Jiawei Dong, Yuchen Song, Jin Li, Min Zhang, Danshi Wang, *Senior Member, IEEE*

***Abstract*—As the key component that facilitates long-haul transmission in optical fiber communications by increasing capacity and reducing costs, accurate characterization and gain settings of erbium-doped fiber amplifiers (EDFAs) are essential for quality of transmission estimation and system configuration optimization. However, it is difficult to construct accurate and reliable EDFA models due to complex physical mechanisms and dynamic loading conditions. Although some mathematical and data-driven models have been proposed, their practical applications will face limitations of intricate parameter measurements and high data requirements, respectively. To overcome limitations of both methods, a physics-informed machine learning (PIML) method for parameter identification and gain estimation of EDFA is proposed, which greatly reduces the data requirements by embedding physical prior knowledge in the neural network. In this approach, the gain of EDFA can be accurately estimated by a physics-informed neural network (PINN)-based forward model when parameters including absorption, gain, saturation, and background loss are known. For practical scenarios where parameters are unknown, PINN-based inverse models are established first to identify actual values of parameters from only several sets of input-output data pairs, and PINN-based forward models are accordingly established for gain estimation with identified values. Moreover, an experimental system is constructed to verify the feasibility and performance of proposed method in practical scenarios. Results show that PIML-based method can effectively identify physical parameters from measured data, and better gain estimation results are achieved with mean absolute error of 0.127 dB and standard deviation of 0.065 dB using identified values than typical values of parameters.**

## I. INTRODUCTION

THE advent of optical amplifier eliminates the need for optoelectronic regenerators, which previously required converting optical signals into the electric domain before regeneration. Optical amplifiers facilitate long-haul transmission in optical fiber communications by increasing capacity and reducing costs. Most systems employ lumped erbium-doped fiber amplifiers (EDFAs) in which losses accumulated over 60 to 100 km of fiber lengths are compensated using short lengths of erbium-doped fibers, and the optical signal is amplified through population inversion of erbium ions. The accurate description of EDFA is essential for its induced-noise characterization for quality of transmission estimation, and its adjustment of gain setting is important for system configuration optimization in optical fiber communications [1]. However, the gain profile of an EDFA is governed by a set of coupled equations of its operating conditions, including channel loadings, control settings, and EDF's own physical features [2], which poses great challenges for the gain modeling of EDFA. Moreover, as optical communications are evolving towards a more flexible and dynamic mode, the more frequently changing loading conditions increases the difficulty of constructing accurate EDFA gain models [3]. Therefore, accurate and reliable gain modeling of EDFA has always been highly valuable yet challenging in optical communications.

To effectively and easily construct a gain model for EDFA, a center of mass (CM) model was proposed [4], which provided a simple mathematical expression for characterizing the gain of EDFAs under different channel loadings. Although few measured data were required, the accuracy of CM model was relatively low. Moreover, the Giles model derived from the particle-level transition theory was proposed for modeling EDFA [5], which mathematically characterized the core dynamic process of pump and signal propagation in the EDF with a set of coupled propagation and rate equations. The key to implementing Giles model lies in the precise identification of numerous physical parameters including absorption coefficients, gain coefficients, saturation parameters, and background loss coefficients. When these physical parameters are determined, the Giles model can be solved numerically. However, the actual values of these parameters are hard to be obtained, and only typical values can be obtained from product manual or specific documents, which may deviate from the actual values due to the variations in production and EDF aging. To obtain the actual values, the intricate measurements are usually required, such as the cutback method and on–off gain method [6], which introduces additional hardware costs, as well as uncertainty-induced inaccuracies and measurement errors.

In addition to the mathematical models, machine learning (ML)-based models have also been employed for the gain modeling of EDFA in a data-driven manner. Without tedious measurements of physical parameters and intricate physical and mathematical knowledge, the mapping relationship between the

This work was supported in part by the National Natural Science Foundation of China under Grant 62171053, Beijing Nova Program under Grant 20230484331, and BUPT Excellent Ph.D. Students Foundation under Grant CX2023226. (*Corresponding author: Danshi Wang*).

Xiaotian Jiang, Jiawei Dong, Yuchen Song, Jin Li, Min Zhang, and Danshi Wang are with the State Key Laboratory of Information Photonics and Optical Communications, Beijing University of Posts and Telecommunications, Beijing 100876, China (e-mail: jxt@bupt.edu.cn; jiaweidong@bupt.edu.cn; songyc@bupt.edu.cn; jinli@bupt.edu.cn; mzhang@bupt.edu.cn; danshi_wang@bupt.edu.cn).

gain profile and operating conditions can be easily established by learning multiple data pairs of input and output power profiles with neural networks [7-9]. Despite the simplicity in implementation, these data-driven ML methods still necessitated a significant volume of data for training, which is also time-consuming and cost-prohibitive in data acquisition. Particularly, acquiring a comprehensive set of gain profiles under different gain settings and channel loading conditions from a single EDFA can take upwards to hundreds of hours [10]. In addition, without considering the underlying physics, these data-driven models may fail to capture the consistent physical behaviors across different EDFA devices, leading to poor generalization performance [11]. Although auxiliary neural networks [12] and transfer learning [13] have been proposed to reduce the data requirements to a certain degree, the established data-driven model remains an opaque "black box" with limited interpretability.

The integration of physical prior knowledge into ML-based models has garnered increasing attentions and been gradually investigated as a promising approach to enhance model accuracy and reduce data dependence. Based on the simplified Giles model, the equivalent flat gain model was developed [14], employing minimal data for linear regression to handle non-flat power fluctuations. In addition, the prior knowledge of CM model was merged into a data-driven neural network [15], moderately decreasing the volume of training data and improving model precision. Additionally, the EDFA gain spectrum was formulated as a simple single-variable linear function derived from physical principles, which considerably diminished the data requirements and maintained consistency across multiple EDFA devices [16]. These methods increase the interpretability of ML models and reduce the dependence on data to some extent. However, a large number of simplifications and approximations are conducted in these hybrid methods, which will make them invalid in some special scenarios. Therefore, how to better combine physical prior knowledge of EDFA with ML to achieve a more comprehensive and practical hybrid EDFA model remains a substantial challenge.

Recently, a physics-informed machine learning (PIML) technology represented by physics-informed neural network (PINN) was proposed [17,18], which combined the benefits of both physical prior knowledge and ML algorithms. It has been verified in multiple fields, including fluid mechanics [19], aerodynamics [20], and fiber optics [21]. By embedding the physical prior knowledge, such as the differential equations and physical constraints (e.g., initial and boundary conditions) of the dynamic system into the loss function, PINN is able to solve the forward problem of learning the system dynamics without training data [22]. Moreover, another potential application of PINN lies in the inverse problem of identifying the equation parameters from the limited measured data, and has accurately identified the equation parameters in fiber optics, such as the nonlinear Schrödinger equation [23] and the stimulated Raman scattering equation [24]. Specifically, PINN continuously learns the dynamics implicit in the data by adjusting equation parameters to fit the measured data, and ultimately making the learned dynamics consistent with the data. Inspired by the solving capabilities of PINN in both forward and inverse problems, it is expected to perform the parameter identification and gain estimation of EDFA. However, in contrast to the general inverse problem, only input and output data can be measured in EDFA, while multiple parameters are wavelength-dependent, which greatly increases the difficulty of parameter identification and gain estimation of EDFA using PINN.

In this paper, a PIML-based method for parameter identification and gain estimation of EDFA is proposed without intricate measurements nor large amount of data. In this approach, a PINN-based forward model can be established directly for gain estimation when physical parameters are known. For practical scenarios where physical parameters are uncertain, a PINN-based inverse model is first established to identify the parameters at each wavelength with only five sets of input and output data pairs. The identified parameters are then adopted as the parameters of differential equations in the PINN-based forward model to estimate the gain of EDFA. Both simulation and experimental verifications are conducted. The results show that all physical parameters can be effectively identified from several sets of measured data, and the estimated gain of EDFA using these identified values is superior to using typical values of the physical parameters. In addition, the settings of inverse models and future works towards more complex scenarios are also discussed. Our work illustrates that as a simple approach without complex measurements, PIML promises seamless migration to different amplifiers with different doping elements and doping concentrations.

## II. Principle

### A. Analytical Model of EDFA

For a fiber amplifier pumped in the 980 nm absorption bands, its dynamics can be characterized by the propagation and rate equations of a homogeneously broadened two-level systems named the Giles model, as shown in Fig. 1(a). Rate equations describe the effects of absorption, stimulated emission, and spontaneous emission on the populations of the ground and metastable states. For the optical signal in the $k^{\text{th}}$ channel, the rate equation under the dynamic channel loadings can be expressed as [5]:

$$\begin{aligned}\frac{dN_2(z,t)}{dt} &= \sum_k \frac{P_k(z,t)\alpha_k}{h\nu_k\zeta\tau} - \frac{N_2(z,t)}{\tau} \\ &\quad - N_2(z,t)\sum_k \frac{P_k(z,t)\left(\alpha_k + g_k^*\right)}{h\nu_k\zeta\tau}\end{aligned} \tag{1}$$

where $N_2(z,t)=\dfrac{\bar{n}_2(z,t)}{\bar{n}_t}$ is the fraction of ions in the excited state. $P_k(z,t)$, $\alpha_k$, and $g_k^*$ are the signal power at distance $z$ and time $t$, wavelength-dependent absorption parameter, and gain parameter in $k^{\text{th}}$ channel; $h$ is Planck's constant; $\nu_k$ is the propagation frequency of the $k^{\text{th}}$ signal; $\zeta = \pi b_{\text{eff}}^2 \bar{n}_2 / \tau$ is the saturation coefficient, where $b_{\text{eff}}$ is the effective doped radius that is equivalent to the core radius in a uniformly doped fiber; $\tau$ is the upper state lifetime of the ions. When the erbium ions are well confined to the center of the optical modes and the EDF is in a steady state, the Giles model will be time-independent. Accordingly, $\frac{dN_2}{dt}=0$ and $N_2$ depends only on $z$ and $P_k$, expressed as:

$$N_2=\frac{\bar{n}_2}{\bar{n}_t}=\frac{\sum_k \frac{P_k(z)\alpha_k}{h\nu_k\zeta}}{1+\sum_k \frac{P_k(z)\left(\alpha_k+g_k^*\right)}{h\nu_k\zeta}} \tag{2}$$

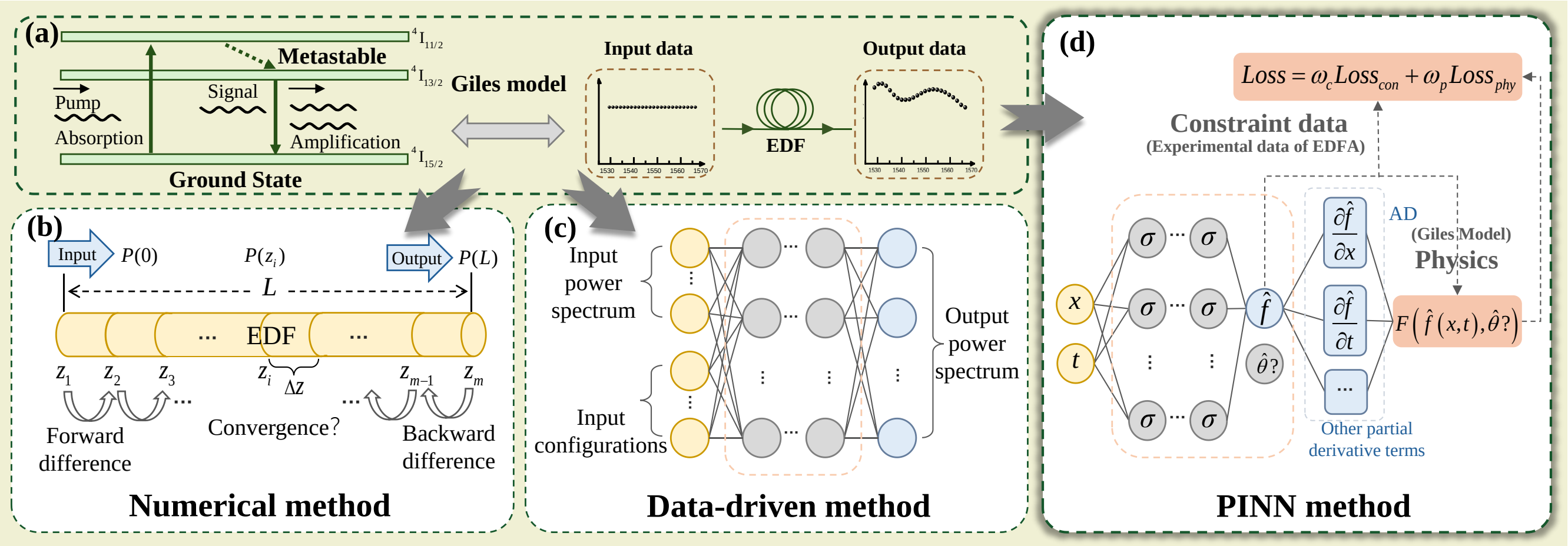

Fig. 1. (a) Analytical Model of EDFA. Schematic of (b) numerical method, (c) data-driven method, and (d) PINN method for EDFA modeling.

At this time, the Giles model is simplified to the propagation equations of pump and signal, which is a set of coupled nonlinear ordinary differential equations (ODEs) at different wavelengths, and the ODE of the signal at $k^{\text{th}}$ channel can be expressed as:

$$\begin{aligned}\frac{dP_k}{dz}&=u_k\left(\alpha_k+g_k^*\right)N_2P_k(z)\\&+u_kg_k^*N_2mh\nu_k\Delta\nu_k-u_k\left(\alpha_k+l_k\right)P_k\end{aligned} \tag{3}$$

In the right-hand side of Eq. (3), the first term represents the gain term, the second term represents the ASE noise term, and the third term represents the absorption term. $u_k$ and $\Delta\nu_k$ are the propagation direction, and frequency bandwidth of the $k^{\text{th}}$ signal; $m$ is the number of polarization modes; $l_k$ is the wavelength-dependent background loss coefficients of the $k^{\text{th}}$ signal. Generally, $l_k$ is assumed to be a constant (i.e., $l_k = l$) within the limited bandwidth of simulated wavelengths [25]. In the ODE of pump, the pump gain coefficient $g_k^*=0$, and pump absorption coefficient $\alpha_p$ is the only unknown wavelength-dependent parameter. When the unknown system parameters set $\theta = [\zeta, l, \alpha_p, \alpha_k, g_k^*]$ are determined, the gain characteristic of the EDF can be accordingly determined.

*B. Numerical Methods for EDFA Modeling*

Generally, the Giles model do not have analytical solutions, and numerical methods such as the finite difference method (FEM) are required. The process of employing the FEM to solve this set of equations is shown in Fig. 1(b). Specifically, it is necessary to solve a boundary value problem consisting of $3n$+1 coupled equations, (i.e., $n$ signal equations, one pump equation, $n$ forward ASE equations, and $n$ backward ASE equations) [26]. The EDF is discretized into several segments from 0 to $L$, with each segment spaced by $\Delta z$. Using the boundary condition data sets of $P(0)$ and $P(L)$, including the signal, pump, and ASE powers at $z = 0$ and $z = L$, and both forward and backward difference formulas, each equation is iteratively solved based on the power state from the previous iteration. This process determines the power set of $P(z_i)$ at each distance $z_i$, including the power of the signal, pump, and both forward and backward ASE. Considering solving form both forward and backward directions, the numerical system requires continuous iteration and examination of the solution's convergence until all the boundary conditions of the equation are satisfied.

It is noteworthy that the choice of $\Delta z$ significantly influences the accuracy and stability of the solution. Excessively large step sizes may lead to unstable or inaccurate results, while excessively small step sizes can result in prohibitively high computational costs. Moreover, numerical methods will fail when $\theta$ are unknown, which limits the practical application of numerical methods.

*C. Data-driven Methods for EDFA Modeling*

Generally, $\theta$ is usually unknown in practical scenarios, and intricate measurements as well as expert knowledge are required. In contrast, only input-output data pairs are required in data-driven methods to establish the gain model for EDFA, and a typical data-driven model for EDFA modeling is shown in Fig. 1(c). The input layer consists of the input power spectrum and the input configurations. The model's output is the predicted power spectrum corresponding to the model's input. By continuously optimizing the loss function composed of the mean square error (MSE) between the predicted results and corresponding labels, the network parameters are accordingly updated. When the loss function eventually converges, the data-driven model can well characterize the physical features of EDFA.

In order for the model to fully learn the dynamic characteristics of the EDFA, it is necessary to gather extensive input-output data pairs under various conditions, including different channel loadings, input and output powers of pump and signals, which places high demands on the quantity and quality of data. Moreover, the underlying physics are usually not considered in data-driven methods, which makes the data-driven models become uninterpretable black box models, and difficult to generalize to different EDFAs without extra data for training.

### *D. PINN Methods for Forward and Inverse Problems*

To overcome the limitations of both methods, the PINN combines machine learning with physical prior knowledge, and serves as a promising method for both forward problem of learning system dynamics and inverse problem of identifying system parameters, without the need for extensive training data. Generally, the dynamic process of a physical system can be mathematically characterized by the differential equations, and a general form of differential equation can be represented as:

$$\frac{\partial f(x,t)}{\partial t} = F(f(x,t),\theta) \quad t \in [0,T], x \in \Omega \tag{4}$$

where $f(x, t)$ is the solution, characterizing the state of dynamic system at the spatial and temporal coordinate $(x, t)$. $\Omega$ represents the spatial domain of $x$, $df(x,t)/dt$ is the temporal derivative term. $F$ is an operator, including spatial and temporal derivatives of each order as well as other linear and nonlinear terms. To solve Eq. (4) with PINN, a series of key steps are required, including model structure construction, application of automatic differentiation (AD) [27], definition of loss function, and the loss optimization process, as shown in Fig. 1(d).

In forward problems of solving differential equations, a neural network is constructed firstly, with the input being spatial and temporal coordinates and the output being the solution at corresponding coordinates. Additionally, the loss function of PINN plays a crucial role, consisting of physical loss term $Loss_{\mathrm{phy}}$ and constraint condition loss term $Loss_{\mathrm{con}}$ composed of initial condition (IC) loss term $Loss_{\mathrm{ICs}}$ and boundary condition (BC) loss term $Loss_{\mathrm{BCs}}$, expressed as:

$$\begin{aligned} Loss_{\mathrm{forward}} &= \omega_{\mathrm{c}} Loss_{\mathrm{con}} + \omega_{\mathrm{p}} Loss_{\mathrm{phy}} \\ &= \omega_{\mathrm{c}} \left( Loss_{\mathrm{ICs}} + Loss_{\mathrm{BCs}} \right) + \omega_{\mathrm{p}} Loss_{\mathrm{phy}} \\ &= \frac{\omega_{\mathrm{c}}}{N_0} \sum_{i=1}^{N_0} | \hat{f}(x_0^i, 0) - f(x_0^i, 0) |^2 \\ &+ \frac{\omega_{\mathrm{c}}}{N_b} \sum_{i=1}^{N_b} | \hat{f}(x_b^i, t_b^i) - f(x_b^i, t_b^i) |^2 \\ &+ \frac{\omega_{\mathrm{p}}}{N_{\mathrm{phy}}} \sum_{i=1}^{N_{\mathrm{phy}}} \left| \frac{d\hat{f}(x_{\mathrm{phy}}^i, t_{\mathrm{phy}}^i)}{dt} - F(\hat{f}(x_{\mathrm{phy}}^i, t_{\mathrm{phy}}^i), \theta) \right|^2 \end{aligned} \tag{5}$$

where $\hat{f}$ is the output of NNs, $\left\{\left(x_0^i, 0\right)\right\}_{i=1}^{N_0}$, $\left\{\left(x_b^i, t_b^i\right)\right\}_{i=1}^{N_b}$, and $\left\{\left(x_{\mathrm{phy}}^i, t_{\mathrm{phy}}^i\right)\right\}_{i=1}^{N_{\mathrm{phy}}}$ are the sets of points for calculating $Loss_{\mathrm{ICs}}$, $Loss_{\mathrm{BCs}}$ and $Loss_{\mathrm{phy}}$, with $N_0$, $N_b$, $N_{\mathrm{phy}}$ be the number of points for calculating these losses. In Eq. (5), $Loss_{\mathrm{ICs}}$ and $Loss_{\mathrm{BCs}}$ are act as regularization terms, $\omega_{\mathrm{p}}$ and $\omega_{\mathrm{c}}$ are adopted as the regularization coefficients of $Loss_{\mathrm{phy}}$ and $Loss_{\mathrm{con}}$, balancing the contributions of two terms in the loss function. To calculate $Loss_{\mathrm{phy}}$, AD is applied to accurately calculate the derivatives of the network output relative to their inputs. Through optimizing the loss function with the gradient descent algorithms, the network parameters are accordingly adjusted. It should be noted that only IC and BC data are required, and no other labelled data are required in the forward model.

In the inverse problems of identifying unknown parameters, system parameters are also set as trainable parameters of the network. During each iteration, the system parameters are also iteratively updated. To accurately estimate the unknown system parameters, extra data constraints are required in the inverse model. Note that IC and BC data are not necessary in the inverse model, and they can be included in the data constraints when they are available. Therefore, $Loss_{\mathrm{con}}$ can be rewritten with the data loss term $Loss_{data}$ instead of $Loss_{\mathrm{ICs}}$ and $Loss_{\mathrm{BCs}}$, and the loss function can be expressed as:

$$\begin{aligned} Loss_{\mathrm{inverse}} &= \omega_{\mathrm{p}} Loss_{\mathrm{phy}} + \omega_{\mathrm{c}} Loss_{\mathrm{con}} \\ &= \omega_{\mathrm{p}} Loss_{\mathrm{phy}} + \omega_{\mathrm{c}} Loss_{\mathrm{data}} \\ &= \frac{\omega_{\mathrm{p}}}{N_{\mathrm{phy}}} \sum_{i=1}^{N_{\mathrm{phy}}} \left| \frac{d\hat{f}(x_{\mathrm{phy}}^i, t_{\mathrm{phy}}^i)}{dt} - F(\hat{f}(x_{\mathrm{phy}}^i, t_{\mathrm{phy}}^i), \hat{\theta}) \right|^2 \\ &+ \frac{\omega_{\mathrm{c}}}{N_{\mathrm{data}}} \sum_{i=1}^{N_{\mathrm{data}}} | \hat{f}(x_{\mathrm{data}}^i, t_{\mathrm{data}}^i) - f(x_{\mathrm{data}}^i, t_{\mathrm{data}}^i) |^2 \end{aligned} \tag{6}$$

where $\left\{ f(x_{\mathrm{data}}^i, t_{\mathrm{data}}^i) \right\}_{i=1}^{N_{\mathrm{data}}}$ represents the dataset measured at points set $\left\{ (x_{\mathrm{data}}^i, t_{\mathrm{data}}^i) \right\}_{i=1}^{N_{\mathrm{data}}}$ with $N_{\mathrm{data}}$ be the number of data points, and $\hat{\theta}$ denotes the trainable system parameters. When the loss function converges, the network's output satisfies the measured data and the learned equation, and $\hat{\theta}$ also converge to the actual values. Since the physical prior knowledge is embedded in the training process, the amount of measured data required is much smaller than that of data-driven methods.

### *E. PIML-based Workflow for Parameter Identification and Gain Estimation of EDFA*

To model the EDFA accurately and efficiently, a PIML-based workflow composed of PINN-based inverse model and forward model for parameter identification and gain estimation of EDFA is proposed, leveraging the physical prior knowledge and several sets of measured data is proposed, as shown in Fig. 2(a). When $\theta$ is known, the PINN model offers a viable alternative to numerical methods, enabling the direct use of the PINN-based forward model to predict the output power spectrum without any labeled data. However, $\theta$ is generally unknown in the majority of cases, where numerical methods are inapplicable. To this end, the PINN-based inverse model is established to identify unknown parameters using only a small amount of data. Subsequently, by adopting the identified parameters in the forward model, it becomes possible to predict the output characteristics of the EDFA under any input configuration.

As the first step in building an accurate EDFA model in practical scenarios, a PINN-based inverse model is first established to identify $\theta$ required for the Giles model in Eq. (3), and its structure is shown in Fig. 2(b). Among the parameters in $\theta$, $\alpha_k$ and $g_k^*$ are wavelength-dependent, while $\alpha_{\mathrm{p}}$, $\zeta$, $l$ are constants in the inverse model. To ensure the accuracy of parameter identification, $\alpha_k$ and $g_k^*$ of different wavelengths are identified separately in the PINN-based inverse model. For identifying the parameters at $k^{\text{th}}$ channel, the input layer of PINN only consists of the propagation distance $z$ as $z$ is the only independent variable in the Giles model, which is different from the general structure of PINN shown in Fig. 1(d). Moreover, since only input and output powers of pump and signal can be measured in practical, multiple sets of constraint data are required to identify all unknown parameters more accurately. Specifically, the PINN establishes a mapping between $z$ to $M$ pairs of pump power $P_{\text{p}}(z)$ and signal power $P_k(z)$ under single channel loading at different input pump power $P_{\text{p}}(0)$ and signal power $P_k(0)$ for the $k^{\text{th}}$ channel. Accordingly, the total constraint condition loss term $Loss_{\text{CON}}$ is the sum of $Loss_{\text{con}}$ composed of BC losses $Loss_{\text{bc0}}$ and $Loss_{\text{bc}L}$ at $z$ = 0 and $z$ = $L$ in $M$ cases, where $L$ is the EDF length, expressed as:

$$\begin{aligned} Loss_{\text{CON}} &= \sum_{i}^{M} Loss_{\text{con}} = \sum_{i}^{M}\left(Loss_{\text{bc0}}^{i} + Loss_{\text{bc}L}^{i}\right) \\ &= \sum_{i}^{M}\left(\left|\hat{P}_{\text{p}}^{i}(0) - P_{\text{p}}^{i}(0)\right|^{2} + \left|\hat{P}_{k}^{i}(0) - P_{k}^{i}(0)\right|^{2}\right) \\ &+ \sum_{i}^{M}\left(\left|\hat{P}_{\text{p}}^{i}(L) - P_{\text{p}}^{i}(L)\right|^{2} + \left|\hat{P}_{k}^{i}(L) - P_{k}^{i}(L)\right|^{2}\right) \end{aligned} \tag{7}$$

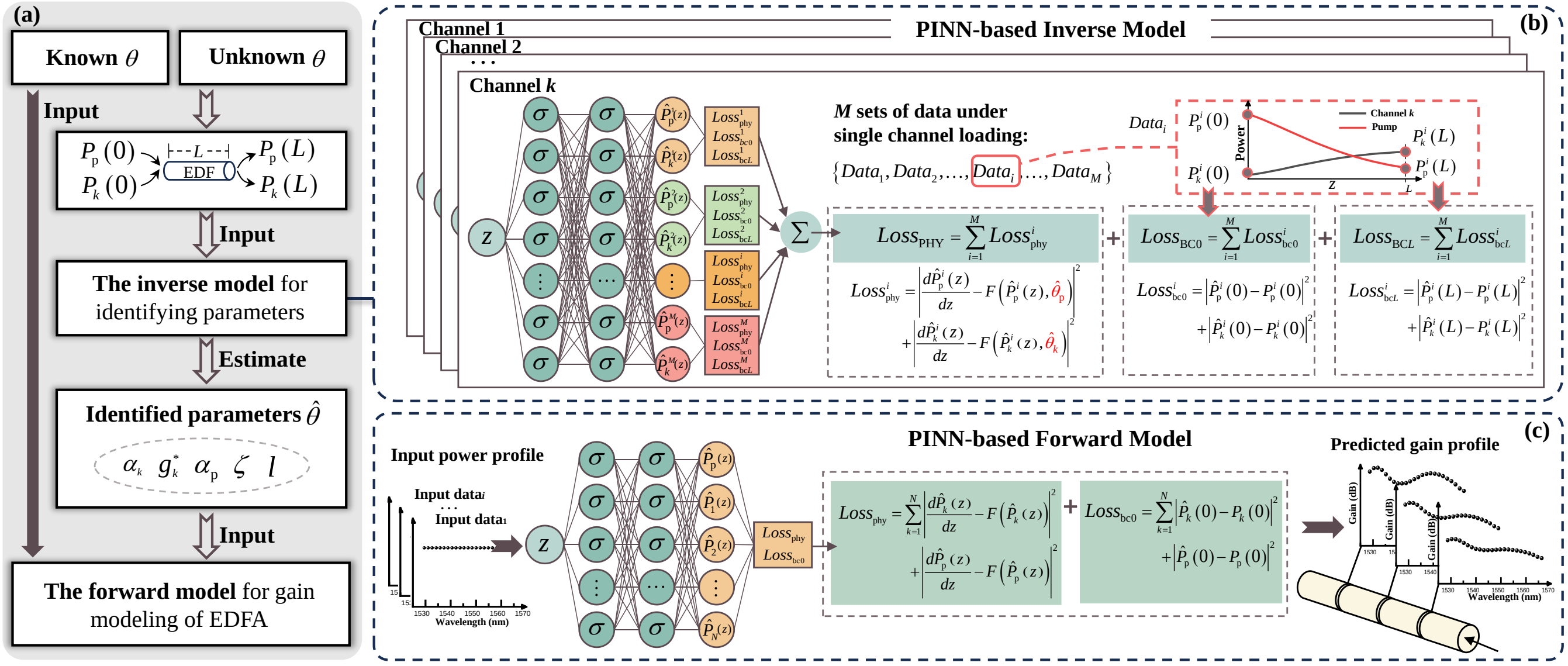

Fig. 2. (a) PIML-based workflow for parameter identification and gain estimation of EDFA, which composed of PINN-based inverse model and forward model. The structure of PINN-based (b) inverse model, and (c) forward model.

In addition to data constraints, the total loss of the $k^{\text{th}}$ inverse model also contains $Loss_{\text{phy}}$ of $M$ cases. Considering the computational efficiency, the simplified form of the Giles model is applied to calculate $Loss_{\text{phy}}$, which ignores the second term that controls the ASE noise, expressed as:

$$\frac{dP_k}{dz} = u_k\left(\alpha_k + g_k^{*}\right)\frac{\bar{n}_2}{\bar{n}_t}P_k(z) - u_k\left(\alpha_k + l\right)P_k \tag{8}$$

Eq. (8) is reasonably accurate when the gain $G \leq 20$dB, because the gain has not been saturated by ASE noise [28]. Here, only the propagations of pump and signal need to be considered, and the equation number is simplified from $3n$+1 to $n$+1, which significantly reduces the number of coupled equations and simplifies the training process. At this time, $Loss_{\text{phy}}$ of each channel only contains two coupled equations at the signal and pump wavelength, and the total physical loss term $Loss_{\text{PHY}}$ can be expressed as:

$$\begin{aligned} Loss_{\text{PHY}} &= \sum_{i}^{M} Loss_{\text{phy}}^{i} \\ &= \sum_{i}^{M}\left(\frac{1}{N_{\text{phy}}}\sum_{j}^{N_{\text{phy}}}\left|\frac{d\hat{P}_{\text{p}}^{i}(z_{\text{phy}}^{j})}{dz} - u_{\text{p}}\left(\alpha_{\text{p}} + 0\right)\frac{\bar{n}_2}{\bar{n}_t}\hat{P}_{\text{p}}^{i}(z_{\text{phy}}^{j}) + u_{\text{p}}\left(\alpha_{\text{p}} + l\right)\hat{P}_{\text{p}}^{i}(z_{\text{phy}}^{j})\right|^{2}\right) \\ &+ \sum_{i}^{M}\left(\frac{1}{N_{\text{phy}}}\sum_{j}^{N_{\text{phy}}}\left|\frac{d\hat{P}_{k}^{i}(z_{\text{phy}}^{j})}{dz} - u_{k}\left(\alpha_{k} + g_{k}^{*}\right)\frac{\bar{n}_2}{\bar{n}_t}\hat{P}_{k}^{i}(z_{\text{phy}}^{j}) + u_{k}\left(\alpha_{k} + l\right)\hat{P}_{k}^{i}(z_{\text{phy}}^{j})\right|^{2}\right) \end{aligned} \tag{9}$$

where $\left\{z_{\text{phy}}^{i}\right\}_{i=1}^{N_{\text{phy}}}$ is the collocation points set for calculating $Loss_{\text{PHY}}$. Accordingly, the total loss of the $k^{\text{th}}$ inverse model can be expressed as:

$$\begin{aligned} Loss_{\text{inverse}} &= \omega_{\text{p}} Loss_{\text{PHY}} + \omega_{\text{c}} Loss_{\text{CON}} \\ &= \sum_{i}^{M}\left(\omega_{\text{p}} Loss_{\text{phy}}^{i} + \omega_{\text{c}} Loss_{\text{bc0}}^{i} + \omega_{\text{c}} Loss_{\text{bc}L}^{i}\right) \end{aligned} \tag{10}$$

Here, only a set of $M$ pairs of input-output pump and signal powers $\left\{\left[\hat{P}_{\text{p}}^{i}(0), \hat{P}_{k}^{i}(0), \hat{P}_{\text{p}}^{i}(L), \hat{P}_{k}^{i}(L)\right]\right\}_{i=1}^{M}$ are required as measured data for estimating unknown parameters $\zeta$, $l$ of the EDF, $\alpha_k$ and $g_k^{*}$ of the $k^{\text{th}}$ channel and $\alpha_{\text{p}}$ of pump. Since the measured data follows the exact equation, PINN is constantly optimizing the unknown parameters until the measured data is well fitted the current equation.

After obtaining all the parameters in $\theta$, the PINN-based forward model can be established for estimating the gain spectrum under different channel loading conditions without additional measured data, as shown in Fig. 2(c). It should be noted that the output of the forward model is no longer the $M$ pairs of pump power and signal power of a single channel, but the powers of all channels and the pump, which is different from the most cases where the outputs of PINN-based forward and inverse models are the same. Therefore, the forward model does not share neurons with the inverse model in this work, and the forward model needs to be trained with $N$ signal powers at $z$ = 0. Specifically, the total loss of the forward model is only composed of $Loss_{\mathrm{phy}}$ and $Loss_{\mathrm{bc0}}$, expressed as:

$$
\begin{aligned}
&Loss_{\mathrm{forward}} = \omega_{\mathrm{p}} Loss_{\mathrm{phy}} + \omega_{\mathrm{c}} Loss_{\mathrm{bc0}} \\
&= \frac{\omega_{\mathrm{p}}}{N_{\mathrm{phy}}} \sum_{i=1}^{N_{\mathrm{phy}}} \left| \frac{d\hat{P}_{\mathrm{p}}(z_{\mathrm{phy}}^{i})}{dz} - u_{\mathrm{p}}\left(\alpha_{\mathrm{p}}+0\right)\frac{\bar{n}_2}{\bar{n}_t}\hat{P}_{\mathrm{p}}(z_{\mathrm{phy}}^{i}) + u_{\mathrm{p}}\left(\alpha_{\mathrm{p}}+l\right)\hat{P}_{\mathrm{p}}(z_{\mathrm{phy}}^{i}) \right|^2 \\
&+ \frac{\omega_{\mathrm{p}}}{N_{\mathrm{phy}}} \sum_{k=1}^{N} \sum_{i=1}^{N_{\mathrm{phy}}} \left| \frac{d\hat{P}_{k}(z_{\mathrm{phy}}^{i})}{dz} - u_{k}\left(\alpha_{k}+g_k^{*}\right)\frac{\bar{n}_2}{\bar{n}_t}\hat{P}_{k}(z_{\mathrm{phy}}^{i}) + u_{k}\left(\alpha_{k}+l\right)\hat{P}_{k}(z_{\mathrm{phy}}^{i}) \right|^2 \\
&+ \omega_{\mathrm{c}} \left| \hat{P}_{\mathrm{p}}(0) - P_{\mathrm{p}}(0) \right|^2 + \omega_{\mathrm{c}} \sum_{k=1}^{N} \left| \hat{P}_{k}(0) - P_{k}(0) \right|^2
\end{aligned} \tag{11}
$$

where $N$ is the number of all loaded channels. Note that $Loss_{\mathrm{phy}}$ includes all $N$ signal equations. Similarly, $Loss_{\mathrm{bc0}}$ is composed of the MSEs of input pump power $P_{\mathrm{p}}(0)$ and input signal power at all $N$ channels (i.e., $P_1(0)$ , $P_2(0)$ ,…, $P_N(0)$ ). When $Loss_{\mathrm{phy}}$ and $Loss_{\mathrm{bc0}}$ are optimized small enough, the power and gain profiles can be obtained. Overall, the proposed PIML-based method effectively addresses both the forward and inverse problems of EDFA with only several sets of data. In this paper, we develop the forward and inverse model based on the repositories of M. Raissi [29] and Sifan Wang [30].

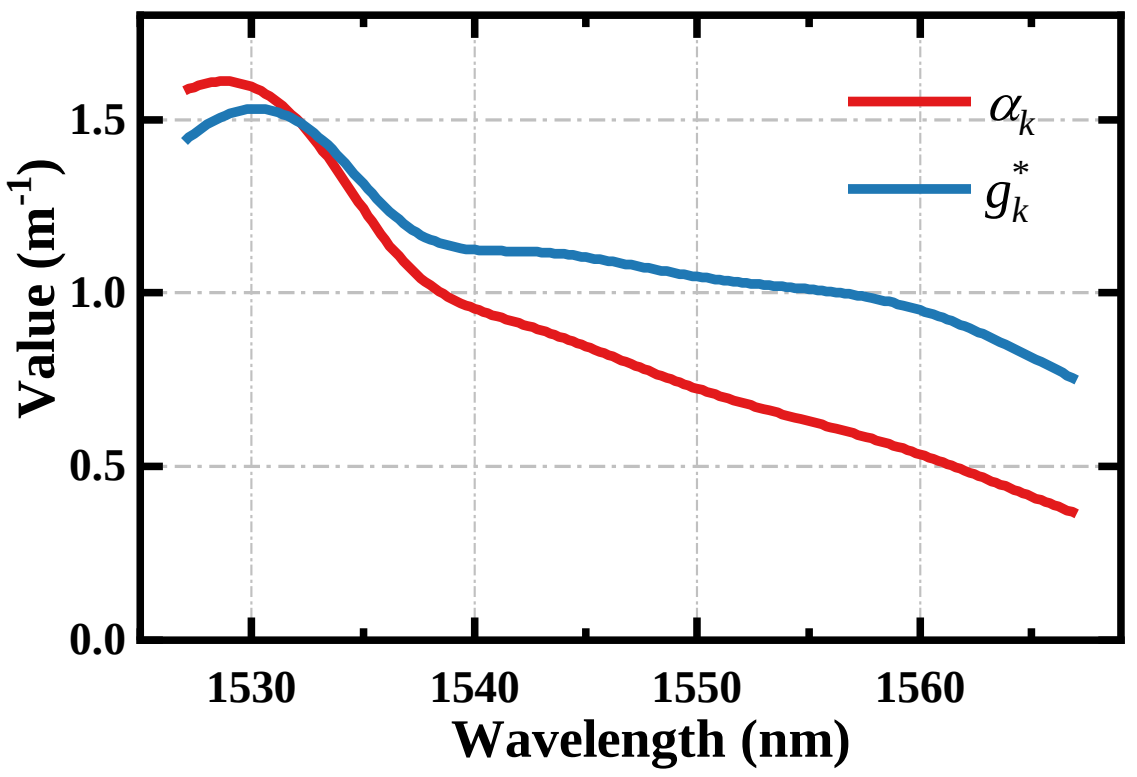

Fig. 3. Absorption and gain coefficients for the EDF used in simulation demonstration of PINN-based forward model and inverse model.

## III. Demonstration of PINN-based Forward Model

When $\theta$ is known, the dynamics of EDFA can be well characterized by the Giles model, and a PINN-based forward model can be applied for the gain estimation of EDFA. In this section, we will validate the performance of the PINN-based forward model on predicting the output gain spectrum under any channel loading condition with known system parameters to illustrate the feasibility and effectiveness of the proposed workflow for this simple scenario.

Here, we consider the signal amplification in an 8 m EDF, where a total of $N_{\mathrm{ch}}$ = 48 C-band transmission channels with the bandwidth of 100 GHz are configured, ranging from 191.4 THz to 196.3 THz. The gain and absorption coefficient curves of the EDF are shown in Fig. 3, while other system parameters are listed in Table I. To learn the system dynamics, a PINN consists of two hidden layers with 200 neurons is constructed, and the ratio of regularization coefficients is set to $\omega_p$:$\omega_c$ = 1:1 in the loss function. To better calculate each term in $Loss_{\mathrm{phy}}$, $N_{\mathrm{phy}}$ = 1024 points are randomly selected from the entire 8 m length of the EDF as input of the model in each iteration. Moreover, an exponential decay learning rate strategy is also implemented in the Adam optimizer to further enhance the stability and convergence speed of the optimization process. Specifically, the initial learning rate was set at $10^{-4}$, decaying exponentially with a decay rate of 0.9 every 5000 steps. This annealing strategy of the learning rate allows for a larger update range in the early stage of training to rapidly decrease the total loss and facilitates fine-tuning of model parameters to ensure convergence to the global optimum in the later stage of training. The MSE of input signal power spectrum is calculated in $Loss_{\mathrm{bc0}}$, and the derivatives in $Loss_{\mathrm{phy}}$ are calculated by AD. After continuous optimization of the loss function, the forward model can be trained to accurately predict the power spectrum at different distances.

TABLE I

Detailed System Parameters of EDF

| Symbol | Parameter | Value |
|---|---|---|
| $l$ | Background loss | 0.02 $m^{-1}$ |
| $b_{eff}$ | Core radius | 1.56 $\mu$m |
| $\bar{n}_t$ | Number density of the dopant ions | $0.955\times10^{25}$ $m^{-3}$ |
| $\tau$ | Upper state lifetime of the ions | 0.01 $s^{-1}$ |
| $m$ | The number of polarization modes | 2 |
| $L$ | The length of EDF | 8 m |

Subsequently, various conditions, including different pump powers, signal powers, and channel load configurations are selected to validate the predictive capability of the PINN-based forward model by comparing with the reference results obtained by the numerical method. Specifically, the estimated gain and absolute error under full-channel (48 channels) loading condition with input power of $P_s$ = -13 dBm per channel are shown in Fig. 4(a) and Fig. 4(b), respectively. It is observed that the gain increases with the pump power, and the predicted values exhibit high consistency with the reference values. The prediction error of the forward model remains below 0.014 dB, and the RMSE is 0.0028 dB. Although the gain error is slightly higher near the 1530 nm channel, the model demonstrates more accurate predictions at longer wavelengths, with a gain prediction error of less than 0.006 dB. Moreover, Fig. 4(c) and Fig. 4(d) show the predicted gain and absolute error when the pump power is 100 mW while all 48 channels are loaded with $P_s$ varying from -13 dBm to -8 dBm. Under this configuration, the error of the PINN model also stays below 0.008 dB, with an RMSE of 0.0025 dB. In addition, the predicted results under different numbers of channel loadings are also investigated in Fig. 4(e), including scenarios with 12, 24, 36, and 48 channels uniformly loaded at a pump power of 100 mW and channel power of -13 dBm. As shown in Fig. 4(f), the model's absolute error is less than 0.015 dB, with an RMSE of 0.0036 dB, and the error at longer wavelengths is around 0.005 dB.

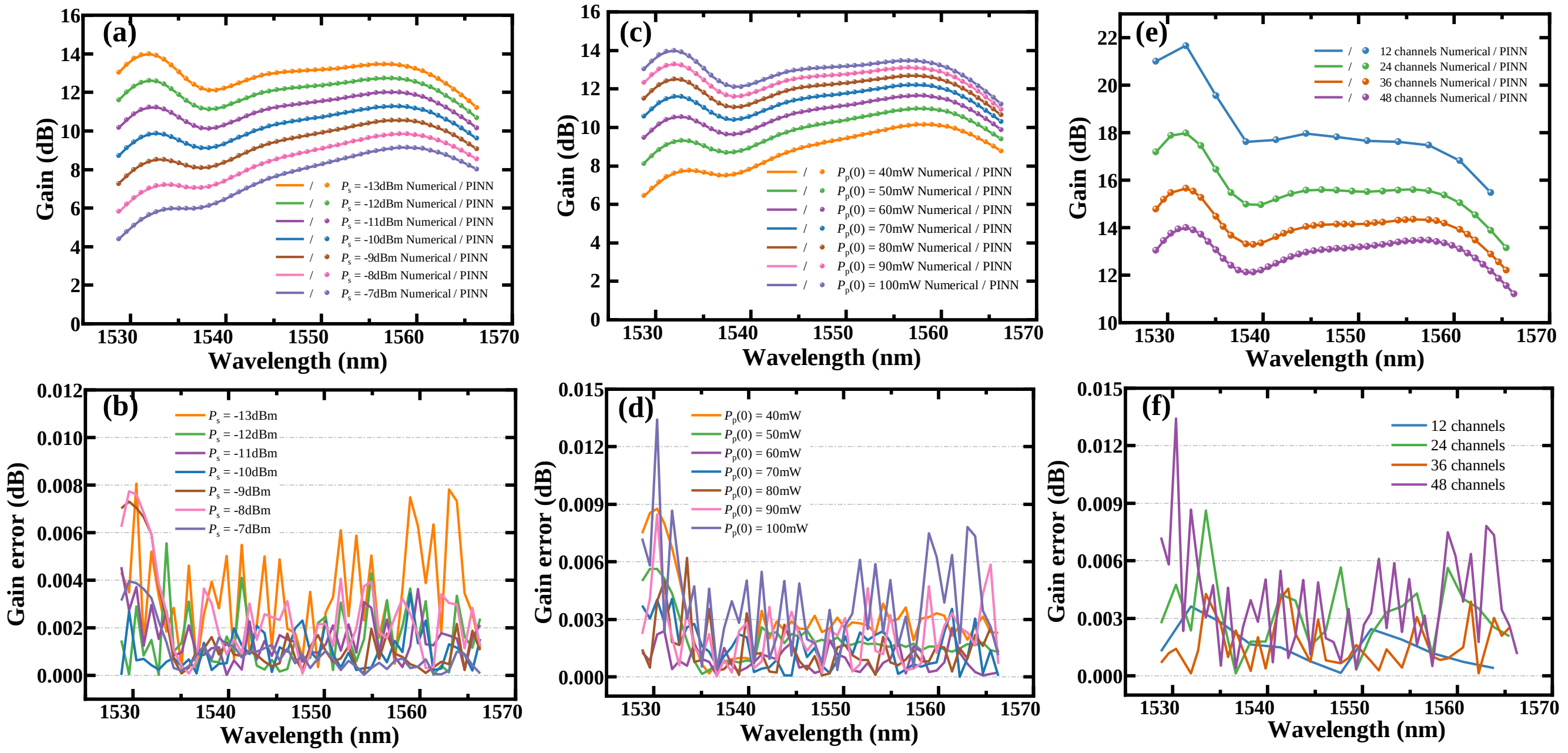

Fig. 4. Results of PINN-based forward model: (a) Gain estimation results and (b) absolute error under different signal powers. (c) Gain estimation results and (d) absolute error under different pump powers. (e) Gain estimation results and (f) absolute error under different number of channels.

## IV. Demonstration of PINN-based Inverse Model

In practical applications, even for EDFs with the same type, their parameters may vary from the typical values due to the fiber aging and differences in the production process. Therefore, $\theta$ is often unknown in most cases, which poses challenges to directly using forward models to predict the output characteristics. To address this issue, a PINN-based inverse model is developed to accurately identify the physical parameters in the Giles model, which is also the key process in the proposed workflow. In this section, the feasibility and performance of the PINN-based inverse model on physical parameters identification will be validated and analyzed.

Here, we consider the parameter identification of the same EDF in Section II, but the fiber radius, the number density of the dopant ions, loss coefficient, absorption coefficient, and gain coefficient are unknown. Accordingly, the physical parameters $\theta$ in the Giles model are required to be accurately first. Since $\alpha_k$ and $g_k^*$ are wavelength-dependent parameters that influence the absorption and gain terms in the equations, different inverse models are established to independently identify them at different wavelengths based on the input-output data pairs of pump and signal under single-channel loading. In addition to $\alpha_k$ and $g_k^*$, wavelength-independent parameters $\zeta$, $l$, and $\alpha_p$ are also identified in each inverse model.

To accurately identify these parameters with the inverse model, data acquisition is essential. However, in contrast to using internal data in most studies on PINN-based inverse models, only the input and output powers of the pump and signals are available, which significantly increases the difficulty of parameter identification. On the one hand, few data constraints make it difficult for unknown parameters to converge to a stable value. On the other hand, the ambiguities of "non-uniqueness issue" in the parameter identification may occur, where unknown parameters converge to incorrect values but the data constraints are well satisfied. Therefore, multiple sets of input-output data pairs under different conditions are required to help parameters converge and eliminate ambiguities caused by "non-uniqueness issue". It should be noted that multiple sets of data will increase the training costs of PINN, and our verification in part A of Section VI shows that five sets of input-output data pairs could achieve a great balance between accuracy and complexity. Therefore, a total of 5 (5 sets of input-output data pairs) × 48 (48 channels in C band) = 240 samples are adopted to calculate $Loss_{\mathrm{CON}}$ in this section. The input-output data for different wavelengths are derived from numerical simulations by solving a boundary value problem with the finite difference method. The calculations for pump and signal power in the simulations are based on Eq. (8), with parameter settings consistent with those described in Section II.

Similar to the structure of the PINN-based forward model, a PINN consisting of two hidden layers with 200 neurons is constructed to identify physical parameters of each channel. $N_{\mathrm{phy}}$ = 1024 points are randomly sampled in the interval of [0, $L$] to calculate $Loss_{\mathrm{PHY}}$ for $k^{\mathrm{th}}$ channel, and the ratio of regularization coefficients is also set as $\omega_p$:$\omega_c$ = 1:1. Under the single-channel loading configuration, $M$ = 5 sets of input and output powers of pump and signal (i.e., $\left\{\left[\hat{P}_{\mathrm{p}}^i(0),\hat{P}_k^i(0),\hat{P}_{\mathrm{p}}^i(L),\hat{P}_k^i(L)\right]\right\}_{i=1}^{5}$) are served as the data constraints. Similarly, the Adam optimizer is applied, with an initial learning rate of $10^{-4}$, decaying to 0.9 of its value every 5000 iterations. According to the prior knowledge about EDFA, $\zeta$ is on the order of $10^{15}$. Therefore, $\zeta$ is scaled to $\zeta' = \zeta / 10^{15}$, and, $\zeta'$ is treated as the unknown parameter to be identified in the inverse model, which helps improve the network's convergence speed, model stability, and identification performance. Moreover, the initial values of the parameters to be identified is also crucial because the model may converge slowly and even incorrectly when the initial values of the parameters are set too far from their actual values. To overcome these limitations, it is better to set the initial values near the typical values of these parameters based on the prior knowledge. Here, it is suitable to set the initial values of $\zeta'$, $\alpha_p$, $\alpha_k$, and $g_k^*$ to 1, and $l$ is initially set to 0 as it is a small value close to 0.

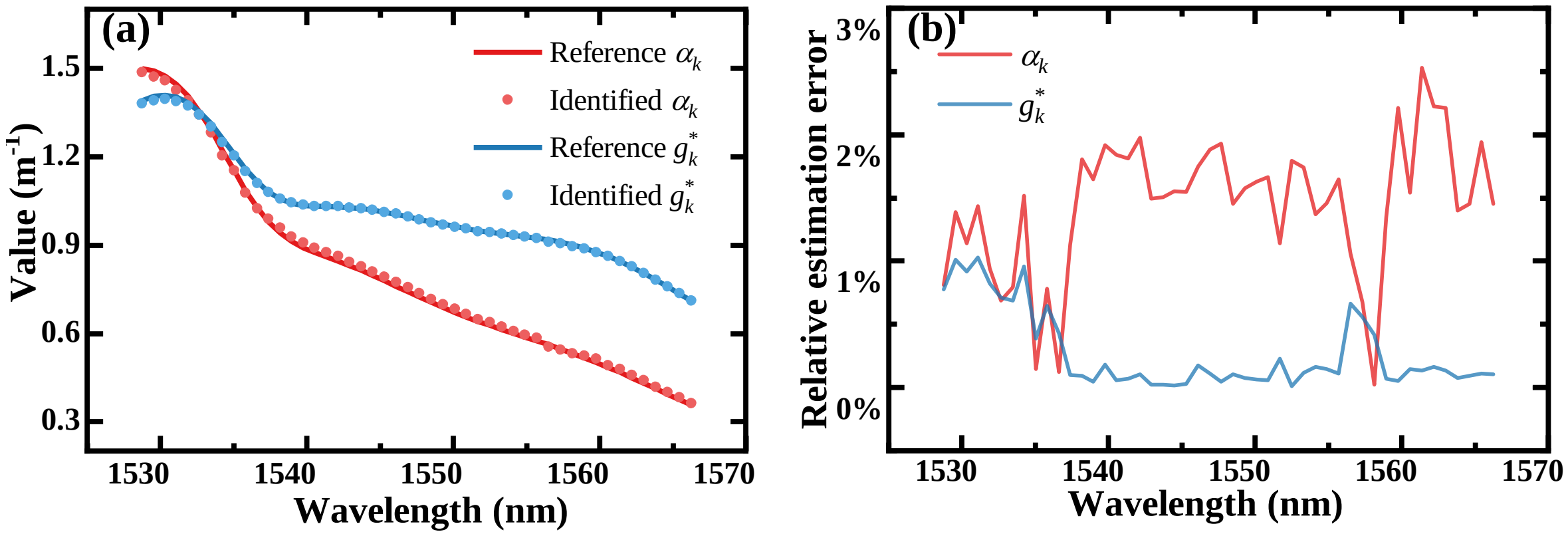

Fig. 5. The identified results of the absorption coefficient $\alpha_k$ and the gain coefficient $g_k^*$ across the C-band. (a) Comparison of identified results with reference values. (b) Relative estimation errors between the identified and reference values

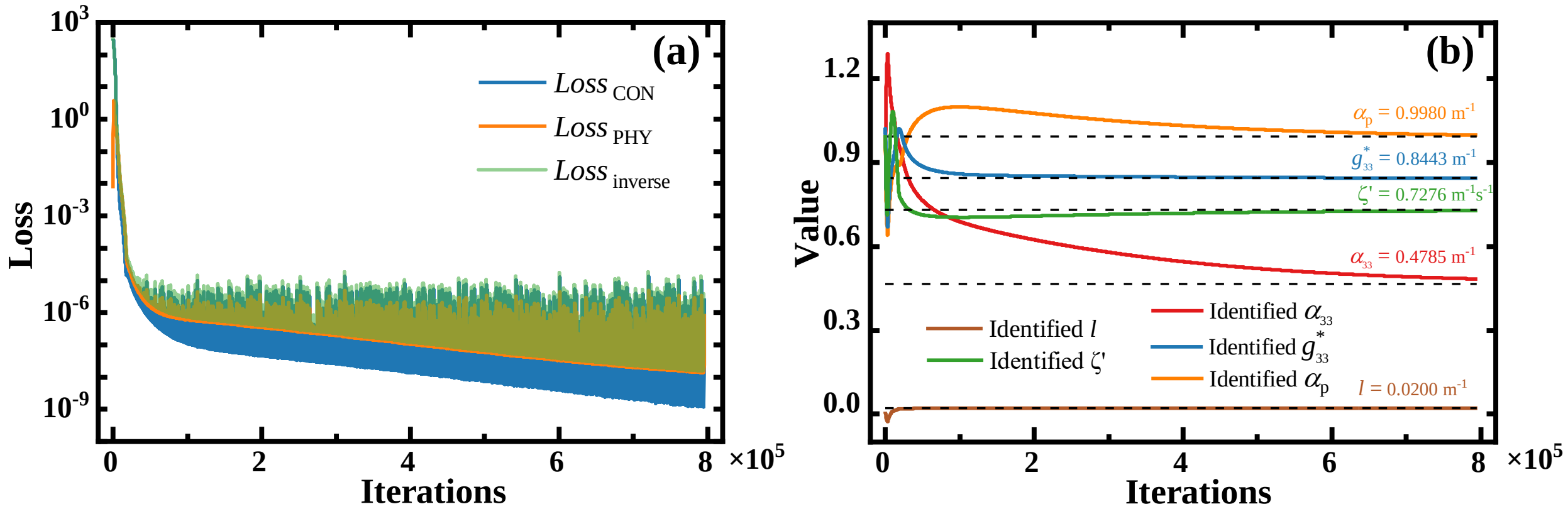

Fig. 6. (a) The loss curve and (b) identification process of all parameters in 33th channel at the wavelength of 1552.52 nm.

As illustrated in Fig. 5(a), the identified values of wavelength-dependent $\alpha_k$, and $g_k^*$ across various wavelengths in the C-band are in good agreement with their reference values. As shown in Fig. 5(b), the average estimation error of $\alpha_k$ is 0.010923, with a maximum relative error not exceeding 2.53%, and the average estimation error of $g_k^*$ is 0.003203, with a maximum relative error not exceeding 1.03%, which illustrates the high accuracy of the PINN-based inverse model. To better illustrate the training process and parameter identification details, the loss curve and identification process are also investigated. Without loss of generality, the loss curve and identification process of all parameters in 33th channel at the wavelength of 1552.52 nm are presented in Fig. 6(a) and Fig. 6(b), which exhibits the highest relative identification error. It can be observed that all loss terms decrease sharply within the first 50,000 iterations, indicating that the model is looking for the right learning direction. Correspondingly, the five parameters to be identified exhibit drastic fluctuations at the early stage of training. However, as the number of iterations increases, all loss terms decrease smoothly, and the five parameters gradually converged to their reference values. After 800,000 iterations, all five parameters converge to their reference values, and the accuracy can be further improved as the loss terms show a continuing downward trend, which demonstrates the potential of the inverse model for higher identification accuracy.

Since the three wavelength-independent parameters $\zeta'$, $\alpha_p$, and $l$ are identified similarly in each inverse model for different channels, the final values of these parameters can be set to the average values of those identified from each inverse model, which is able to reduce the error in a single identification. Here, we identify these values under 50 different sets of boundary conditions and wavelengths, and the statistical results of these three parameters are shown in Fig. 7. It can be intuitively seen that the statistical results of the three parameters approximate Gaussian distribution, and the final identified values are set as their mean values (i.e., $\alpha_p$ = 0.991, $l$ = 0.020, and $\zeta'$ = 0.729), with their respective relative errors being 0.217%, 0.05%, and 0.155%, which is relatively lower than that of the wavelength-dependent parameters. Notably, the relative errors for these three parameters are consistently maintained within 0.824%, 0.2%, and 0.977% in each inverse model, proving the effectiveness and accuracy of the average strategy.

Furthermore, when the three wavelength-independent parameters are determined, only $\alpha_k$ and $g_k^*$ for each channel are required to be identified, which reduces the number of parameters to be identified and significantly improves the convergence speed. Therefore, the three parameters $\alpha_p$, $l$, and $\zeta'$ can be first identified by several inverse models. Next, these parameters can be applied as the known parameters in the inverse models for other channels, and only $\alpha_k$ and $g_k^*$ need to be identified. As shown in Fig. 8, when the average values of the three wavelength-independent parameters are adopted as the equation parameters, the parameter identification process of the 33th channel coverages at a faster speed compared to the five-parameter identification process shown in Fig. 6(b). When five parameters are identified, 800,000 iterations are required and the relative errors are 2.527% and 0.132% for $\alpha_k$ and $g_k^*$, respectively. However, only 400,000 iterations are required when only identifying the two wavelength-dependent parameters, with relative errors of 0.928% and 0.049%, respectively.

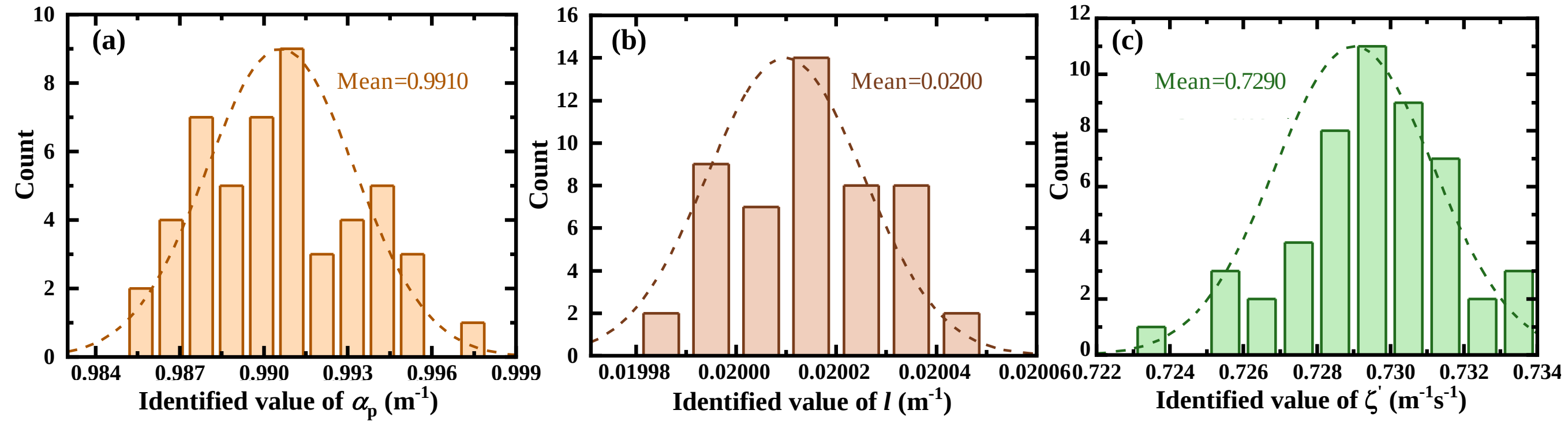

Fig. 7. Statistical results of the three wavelength-independent parameters (a) $\alpha_p$, (b) $l$, and (c)$\zeta'$ under 50 different sets of boundary conditions and wavelengths.

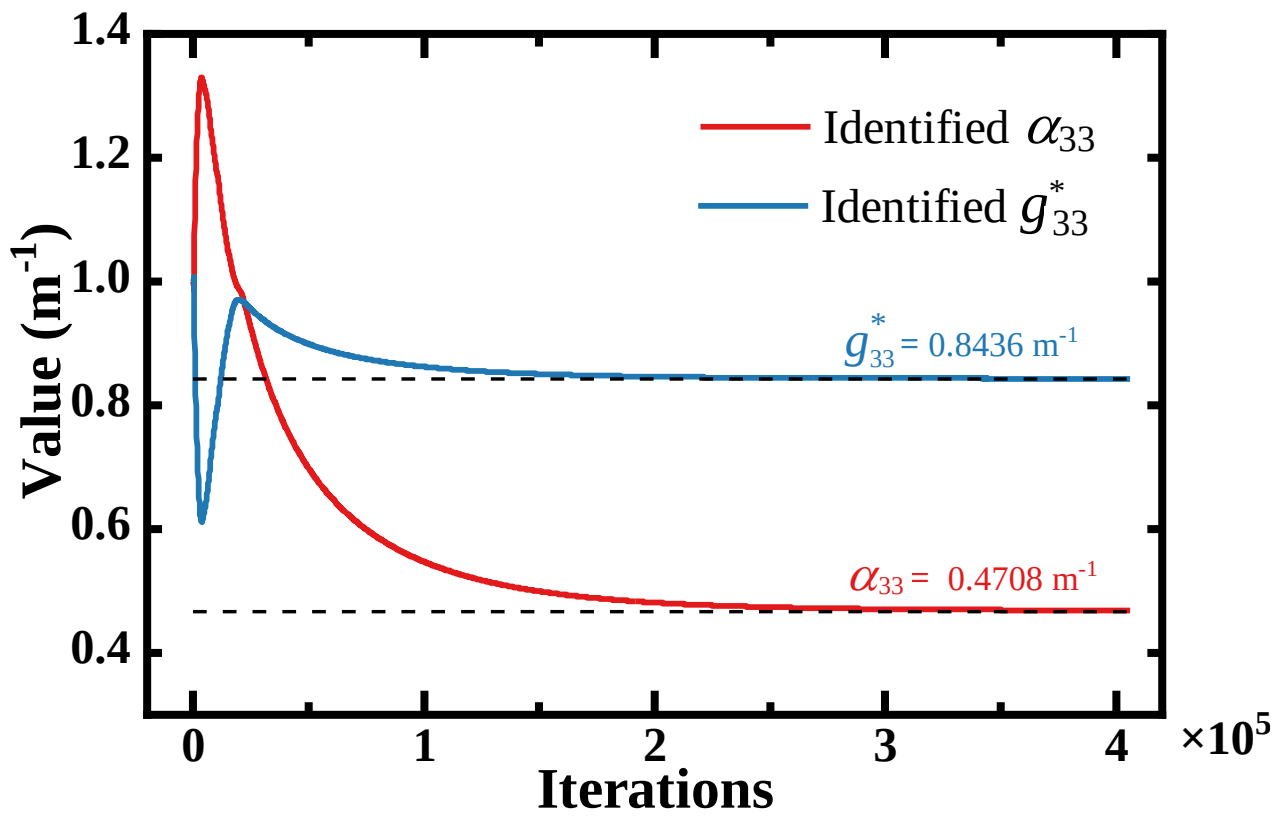

Fig. 8. The identification process of wavelength-dependent parameters in 33th channel when the three wavelength-independent parameters are determined.

When all the unknown parameters are identified, they can be adopted in the PINN-based forward model for gain estimation of EDFA under various channel loading configurations and pump powers. Taking the gain profiles under different pump powers as an example, the comparison and gain errors between the gain profiles using parameters identified and the reference parameters at different pump powers are shown in Fig. 9(a) and Fig. 9(b), respectively. Here, 7 scenarios under different pump powers are tested, and thus a total of 7 × 48 = 336 samples are adopted for gain error comparison. The predicted gain profiles exhibit good consistency with the reference profiles, where the RMSE is only 0.0654 dB, and the maximum gain prediction error does not exceed 0.2 dB. Compared with Fig. 4(c) and Fig. 4(d), the gain error is slightly higher due to the subtle inaccuracies of identified parameters. Similarly, the gain prediction errors are relatively larger in the shorter wavelength channels of the C-band, and stabilizes at below 0.1 dB for wavelengths exceeding 1540 nm.

## V. Experimental Demonstration of PIML for EDFA

In this section, an experimental system for signal amplification is constructed to further verify the feasibility and performance of the proposed PIML-based workflow for parameter identification and gain estimation in practical scenarios. Here, only typical values of few physical parameters are available, and they may deviate from the typical values due to fiber aging and subtle differences in production. Therefore, PINN-based inverse models are first established to identify the actual values of all parameters based on the data measured from the experimental system. Then, a PINN-based forward model is established based on the identified parameters for EDFA gain estimation, and the effectiveness of the workflow is verified by comparing the gain predicted with the gain measured.

### *A. Experimental Setup*

The schematic of experimental system for signal amplification is shown in Fig. 10. First, the pump is generated by a 976 nm pump laser source and coupled with the beam (viewed as the signal) generated by a C-band tunable laser source through a coupler. Next, the output beam of the coupler is input to a $L$ = 8 m EDF from OFS Fitel, LLC, which is a low-doped variety designed for C-band signal amplification. At the same time, an optical spectrum analyzer (OSA) is used to obtain the input powers of pump and signal by measuring the input power spectrum from the wavelength of 900 nm to 1700 nm. Then, the pump and signal are absorbed and amplified when they propagate in the EDF, respectively. Similarly, the output power spectrum is also measured using the OSA to obtain the output powers of pump and signal.

To identify the wavelength-dependent parameters $\alpha_k$ and $g_k^*$, twelve equally wavelength-spaced channels within the C-band are investigated in this study. Similarly, $M$ = 5 different power sets of the signal and pump are set as the inputs of the EDF for each channel, and the detailed power settings are list in Table II. Notably, the selection of the five sets of power values for pump and signal is not unique. To mitigate the impact of ASE noise on the output data, a relatively high signal power is selected to ensure the gain of signal remains below 20 dB. Moreover, the intervals between five input pump powers are set relatively large so that the inverse model can fully learn the amplification characteristics in a broad gain range. Unlike the simulation data, the experimental data may introduce additional errors caused by the precision of instruments, changes in the experimental environment, and inherent noise in optical signals. To further mitigate the ASE noise and effects of these additional errors, five independent measurements are conducted for each power setup in Table II (i.e., 25 sets of measurements are required for Giles parameters identification in each channel), and the average power values are used as the measured data for parameter identification. In this section, Giles parameters at 12 different channels equally spaced in the C-band are identified, and thus a total of 25 × 12 = 300 measurements is performed.

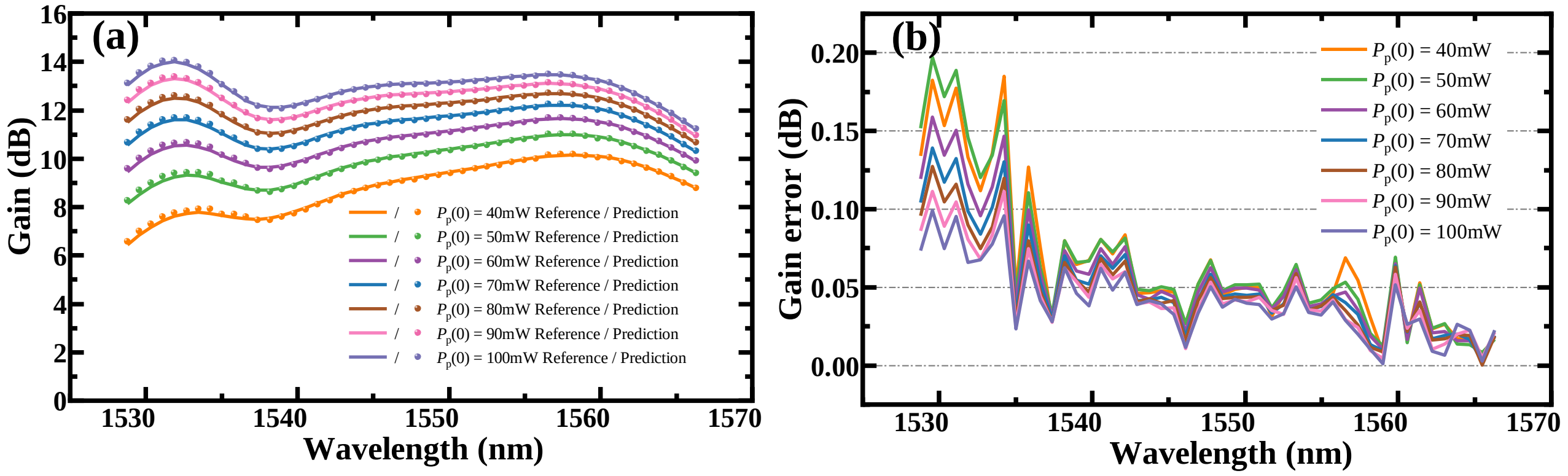

Fig. 9. (a) The gain estimation results of the PINN-based forward model using parameters identified by the inverse model. (b) The gain error between the estimated gain and the reference gain at different pump powers.

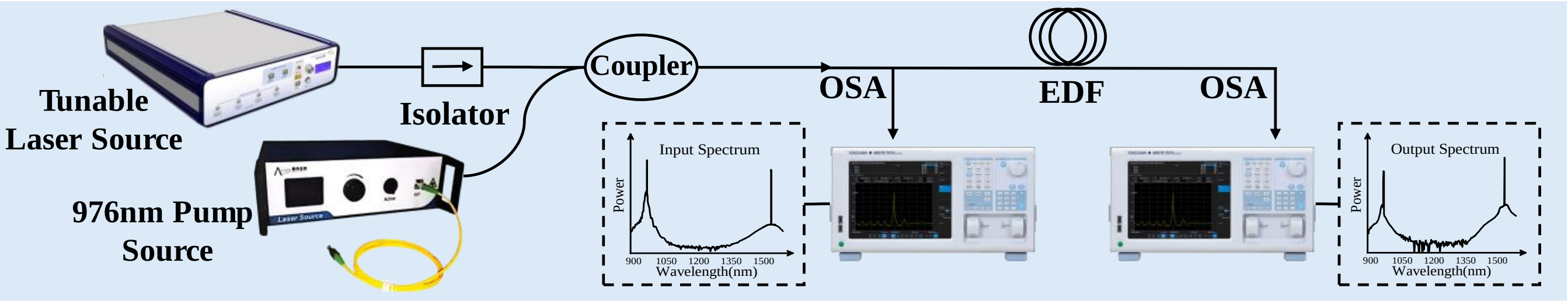

Fig. 10. Schematic of experimental setup for signal amplification under single-channel loading.

TABLE II
POWER SETUP FOR THE SIGNAL AND PUMP LIGHT

| Data List | Signal Power (dBm) | Pump Power (dBm) |
|---|---|---|
| Data 1 | -0.72 | 13.82 |
| Data 2 | 0.28 | 17.17 |
| Data 3 | 1.28 | 18.89 |
| Data 4 | 2.28 | 20.26 |
| Data 5 | 3.28 | 21.31 |

*B. Result of PINN-based Workflow on Experimental Data*

After collecting the input and output power values of pump and signal at different wavelengths, PINN-based inverse models are first established to identify $\theta$ at different wavelengths. Similarly, the structure and hyperparameters of the PINN-based inverse model are consistent with the inverse model for simulation data in Section IV. Since the error in experimental data is unavoidable, overfitting the measured data will lead to large parameter identification errors. Therefore, it is crucial to adjust the regularization coefficient of each loss term to make the model converge to the global optimum. By setting a larger $\omega_p$, the inverse model is encouraged to adhere more closely to physical laws rather than to perfectly fit the boundary condition data. Here, the ratio of regularization coefficients is set to $\omega_p$:$\omega_c$ = 50:1 in the loss function, preventing the model from overfitting the noisy experimental data at the boundaries, which may cause the model to learn incorrect parameter values. Moreover, 800,000 iterations are conducted in each inverse model to make the network converge stably on experimental data.

When all the inverse models eventually coverage, the parameter identification results of wavelength-dependent and wavelength-independent parameters are shown in Fig. 11(a) and Fig. 11(b), respectively. For wavelength-dependent $\alpha_k$ and $g_k^*$, the identified values exhibit similar trends to the typical values of EDF applied. However, large statistical fluctuations occur in Fig. 11(a), which is different from Fig. 5(a). On the one hand, the impact of ASE noise cannot be fully removed. On the other hand, factors such as EDF aging and production process differences will also cause the actual parameter values to deviate from the typical values. Similarly, the three wavelength-independent parameters identified in different inverse models exhibit significant stability and also closely to but not exactly match the corresponding typical values. To better illustrate the learning details, the loss curve and parameter identification process are also investigated. Without loss of generality, the loss curve and parameter identification process of the 8th channel at wavelength of 1552.52 nm are shown in Fig. 12(a) and Fig. 12(b), which exhibits the largest gain prediction error in the subsequent validation of the PINN-based forward model. Similar to Fig. 6(a), each loss term in Fig. 12(a) rapidly decreases in the early stages of training, and it gradually coverages as the iteration increases. Since $Loss_{\text{phy}}$ is given a larger weight, it is about two orders of magnitude lower than $Loss_{BC}$, which illustrates that the inverse model strictly follows the physical laws and avoids overfitting of the data. During the parameter identification process, all parameters show large fluctuations at the beginning, which corresponds to the network looking for clear learning directions in the early stage of training. With the iterations increase, all parameters are optimized in the direction of making the loss function smaller and eventually converage.

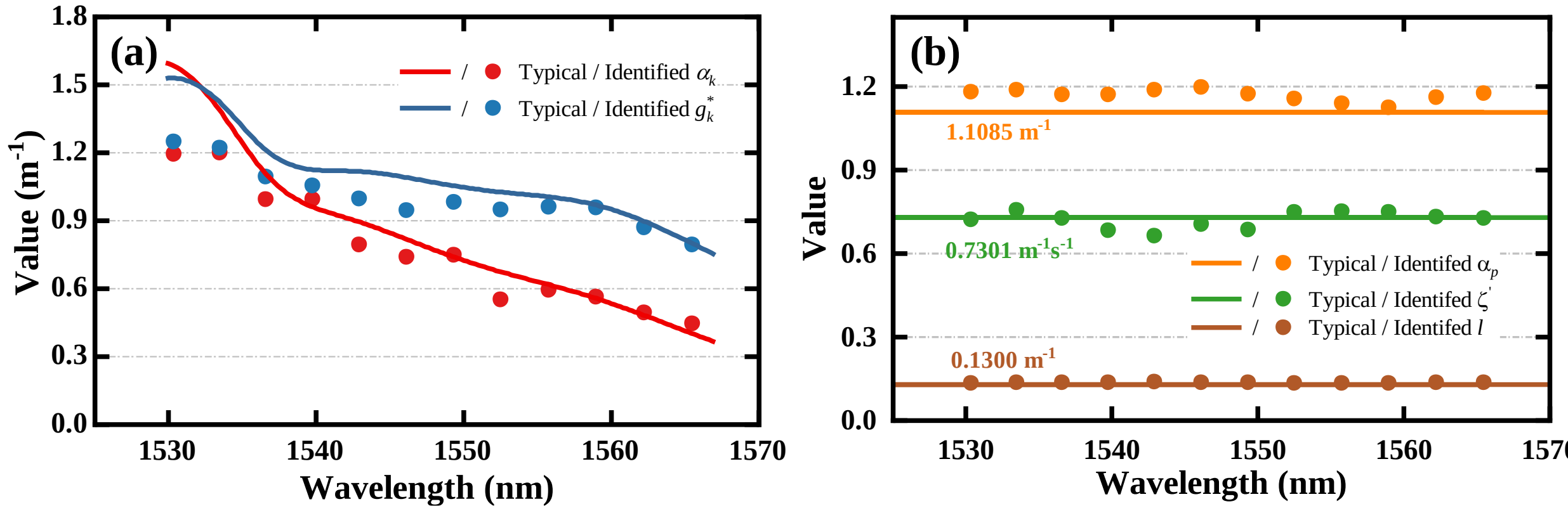

Fig. 11. Identified results of (a) wavelength-dependent parameters. (b) wavelength-independent parameters.

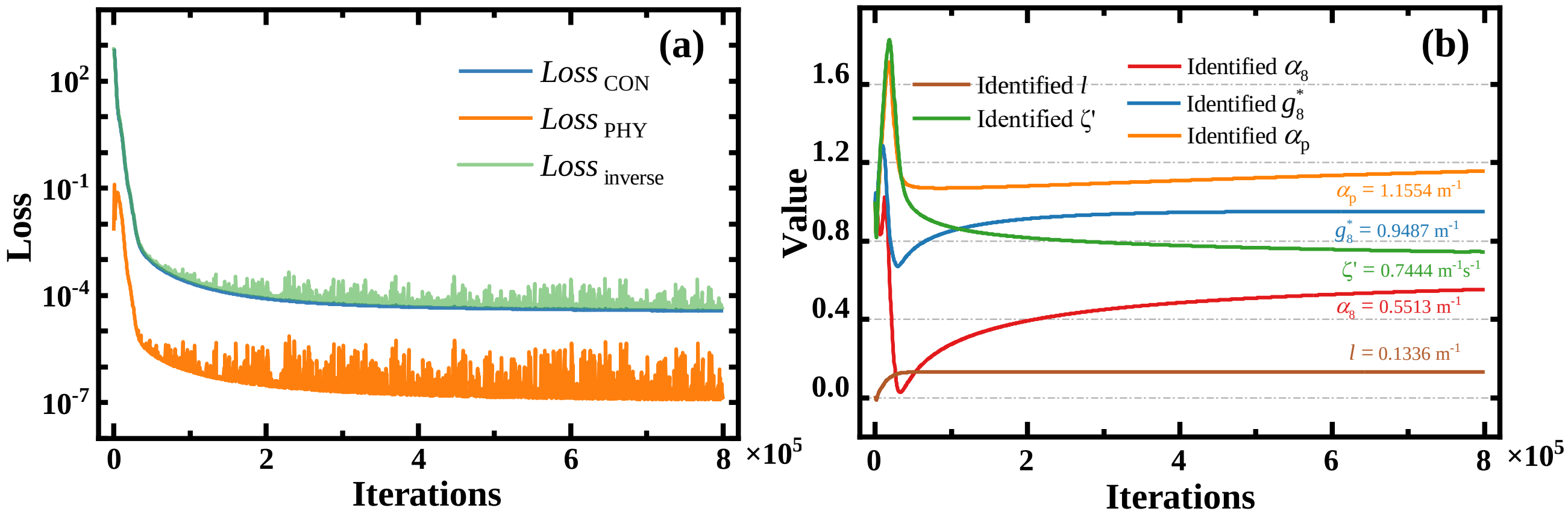

Fig. 12. (a) The loss curve and (b) identification process of all parameters in $8^{th}$ channel at the wavelength of 1552.52 nm.

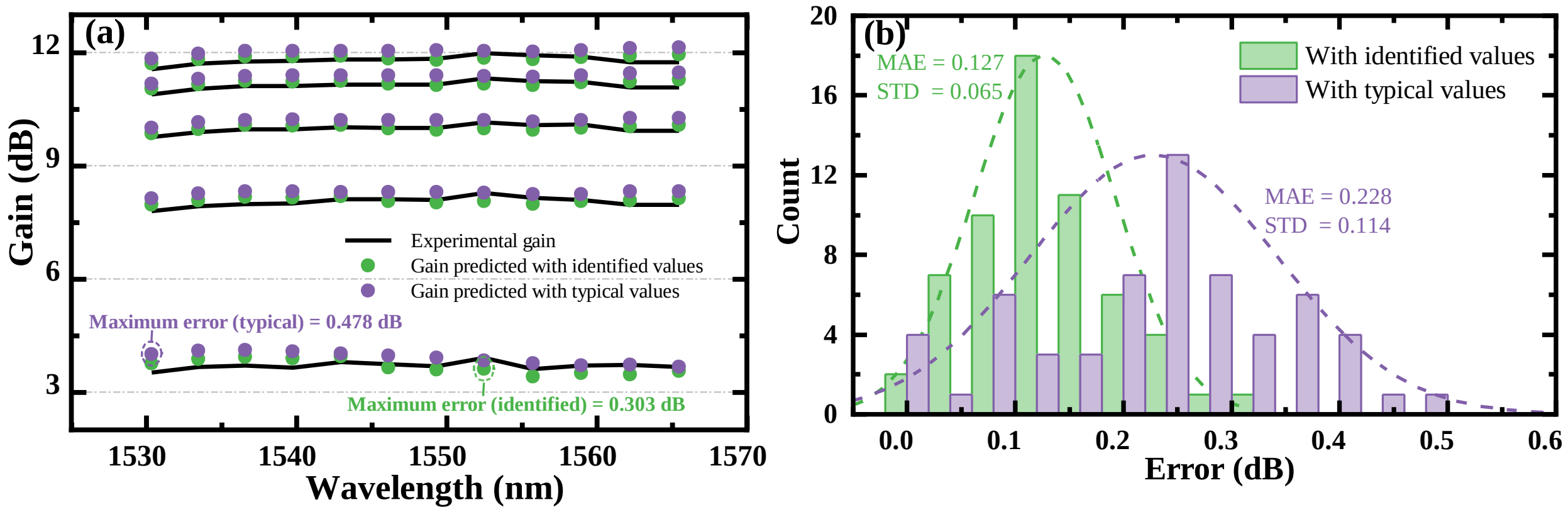

Fig. 13. Results of PINN-based forward model: (a) Gain estimation results and (b) error statistical results using typical values and identified values, respectively.

To verify the accuracy of the parameters identified and construct the gain model of EDFA, a PINN-based forward model is further established. The structure and hyperparameters of the forward model are aligned with those detailed in Section III. Moreover, the identified values of $\alpha_k$ and $g_k^*$ as well as the average values of identified $\alpha_p$, $l$, and $\zeta'$ are applied as the system parameters in the forward model. Here, a total of 5 (5 sets of input-output data pairs) × 12 (12 channels in C band) = 60 measurement samples are adopted as the test samples. As the comparison between experimental gain and estimated gain shown in Fig. 13(a), the gain predicted using identified values are closer to the experimental gain with the maximum error of 0.303 dB. However, the gain estimated using typical values exhibit a larger deviation from the experimental results with the maximum error of 0.478 dB. Particularly, predicted gain using typical values exhibit higher errors in short-wavelength channels, where identified values deviate greatly from the typical values, illustrating the effectiveness of the inverse models. To better illustrate the accuracy and reliability of the forward model with identified values, error statistical results using typical and identified values are shown in Fig. 13(b). When using identified values in the forward model, smaller mean absolute error (MAE) of 0.127 dB and standard deviation (STD) of 0.065 dB are achieved, which are 0.101 dB and 0.049 dB lower than MAE and STD achieved using typical values, demonstrating that the identified parameters are more accurate and the PINN-based forward model is able to effectively and accurately estimating the gain of EDFA.

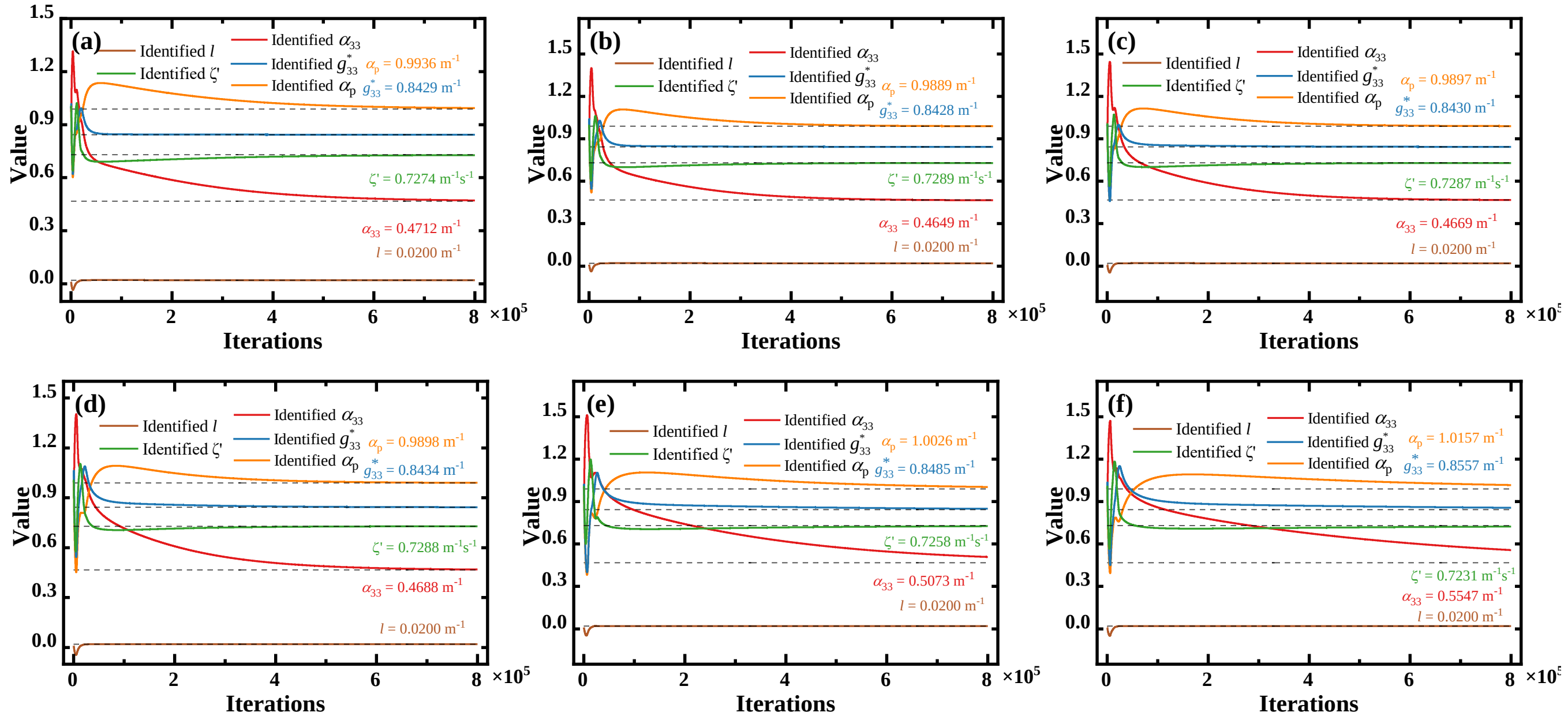

Fig. 14. The identification process of all parameters in 33th channel using the same configuration as Fig. 6(b), except that (a) $M$ = 3, (b) $M$ = 4, (c) $M$ = 5, (d) $M$ = 6, (e) $M$ = 7, (f) $M$ = 8.

It should be noted that for amplifiers with different doping elements and doping concentrations, the typical values of the physical parameters are often partially or even impossible to obtain, which will greatly limit the application of numerical methods. At the same time, data-driven methods will also show poor generalization ability in the new scenarios without a large amount of new data for training. In contrast, the proposed PINN-based workflow can be seamlessly migrated to the new amplifiers with only several sets of new data.

## VI. Further Analysis and Discussion

The previous sections have verified the feasibility and accuracy of PINN-based inverse and forward models through simulation and experimental results. To better cope with practical scenarios, this section will analyze the settings of the inverse model, including the selection of $M$ and the settings of the initial values of unknown parameters. Moreover, future works such as dynamic channel loading scenarios and commercial EDFA modeling are also discussed and prospected.

### *A. Settings in the Inverse Model*

In previous verifications, $M$ = 5 sets of input-output pump and signal power pairs are adopted to calculate the $Loss_{\mathrm{CON}}$ of the inverse model. However, the setting of $M$ is flexible and is essentially a trade-off between model accuracy and complexity. To illustrate this point, the inverse models using $M$ = 3–8 sets of data pairs are tested with the same network settings as Fig. 6. Their training details of 800,000 iterations and relative error of each parameter after network convergence are shown in Fig. 14 and Table III, respectively. In general, at least $M$ = 5 sets of data constraints should be adopted to avoid the unknown parameters converging to other values in the inverse problem of identifying five unknown parameters. Particularly, the models using $M$ = 3 and $M$ = 4 sets of data pairs can also converge in the correct direction under the current network configuration, and thus we also present the results of these two cases to better illustrate the balance between model accuracy and complexity. Under the same network structure, different models show similar parameter learning details, but the model adopting smaller $M$ converges with fewer iterations. When 800,000 iterations are performed, the models adopting $M$ = 6, 7, and 8 have not converged, as shown in Fig. 14(d) - Fig. 14(f). Although it is not strictly followed, the unknown parameters are identified more accurately when more data pairs are used, as the relative errors listed in Table III. Considering that the accuracy improvement is not large, $M$ = 5 sets of data pairs are adopted in this paper for parameters identification, which achieves a good balance between accuracy and complexity.

TABLE III
Relative Errors of Identified Results After Network Converged

| $M$ | $\alpha_k$ | $g_k^*$ | $\alpha_p$ | $l$ | $\zeta'$ |
|---|---|---|---|---|---|
| 3 | 1.018% | 0.043% | 0.480% | 0.050% | 0.218% |
| 4 | 0.155% | 0.048% | **0.005%** | 0.050% | 0.067% |
| 5 | 0.095% | 0.021% | 0.086% | 0.050% | 0.035% |
| 6 | 0.120% | 0.025% | 0.094% | 0.050% | 0.033% |
| 7 | 0.095% | 0.023% | 0.072% | 0.050% | 0.023% |
| 8 | **0.078%** | **0.020%** | 0.055% | 0.050% | **0.020%** |

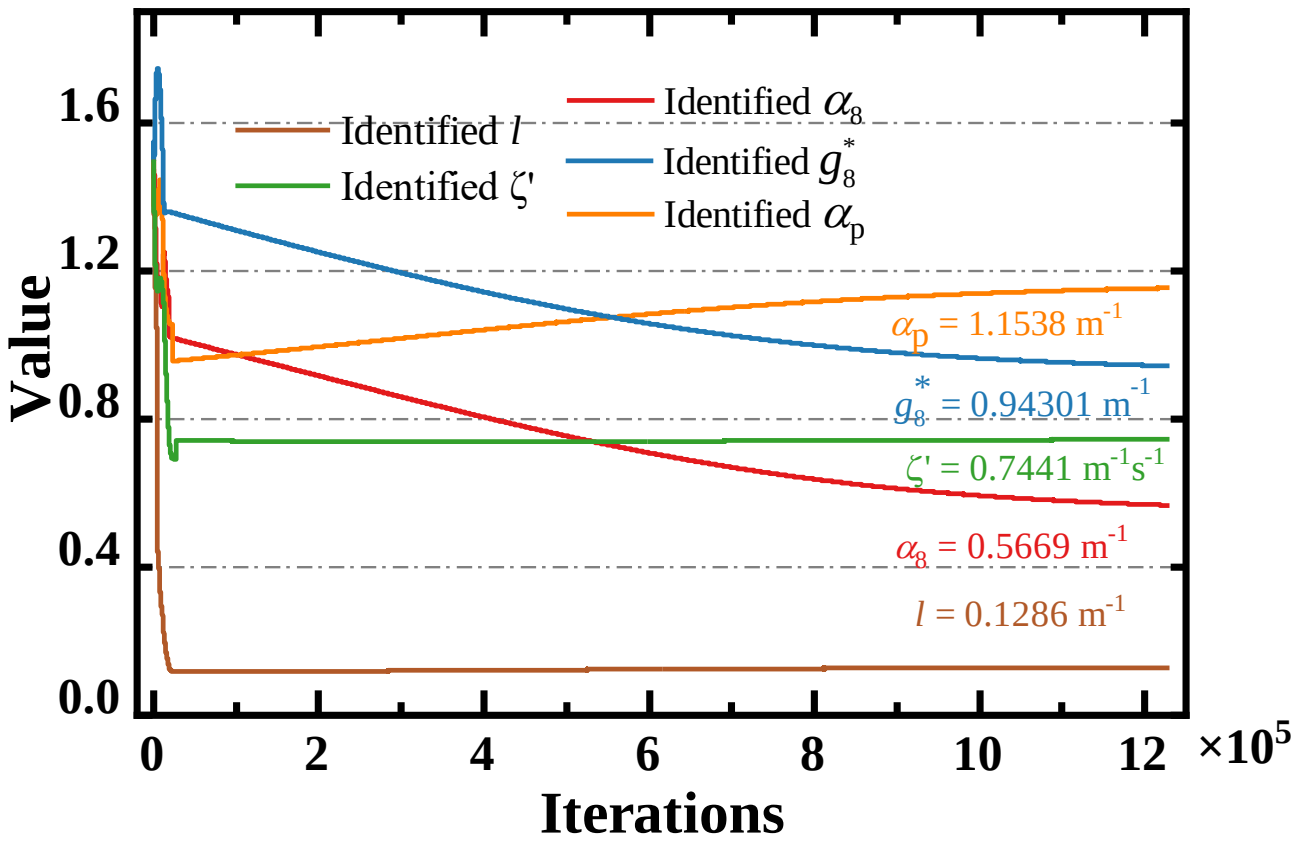

Fig. 15. The identification process of all parameters in 8th channel using the same configuration as Fig. 12(b) of the manuscript, except that the initial values of all parameters are set to 1.5.

Another key setting in the inverse model is the initial values of all unknown Giles parameters, which will affect the accuracy and efficiency of parameters identification. The initial values of most Giles parameters are set to 1 (Except that $l$ is initially set to 0) in previous verifications. Among all the parameters to be identified, the largest difference from the initial value of 1 is the typical value of $\alpha_k$ at 1565 nm (i.e., 0.45 dB/m), which is almost twice of its value. Intuitively, the farther the initial values are from the true values, the more difficulties will be facing in parameter identification. A common strategy for parameter identification with PINN is set the initial values of the unknown parameters to their typical values if possible. In fact, the inverse model is able to accurately identify parameters over a wide range, and Fig. 15 show the identification process of all parameters in 8th channel using the same configuration as Fig. 12(b), except that the initial values of all parameters are set to 1.5. Similarly, all parameters show large fluctuations at the beginning, and all parameters are gradually optimized towards convergence with the iterations increase. However, since the initial value deviates greatly from the actual value, more iterations than Fig. 12(b) are required for optimization. Therefore, setting suitable initial values (e.g., 1 for [$\zeta'$, $\alpha_p$, $\alpha_k$, $g_k^*$], and 0 for $l$) will help the network converge faster, which is significant when evaluating individual EDFs whose typical parameters are unknown.

### *B. Future Works*

In the future, more accurate modeling of EDFA under dynamic loadings and actual devices will be required to realize digital twin of optical networks [32]. By identifying the Giles parameters for the Giles model at steady state in Eq. (3), the subsequently established forward model is able to predict the gain spectrum under any channel loading. However, it does not reproduce the time transients expressed in Eq. (1). For more complicated scenario where channel loading changes dramatically, it is required to jointly solving the coupled equations composed of Eq. (1) and Eq. (3), which is more accurate for dynamic scenarios. To this end, additional variables $t$ and $N_2(z, t)$ need to be added to the input and output layers of the inverse or forward model, respectively. Since $N_2(z, t)$ is hard to measure in practice, it is challenging to identify the Giles parameters from scratch by only measured data pairs of the input and output pump and signal powers. In fact, it is possible to jointly optimize $N_2(z, t)$ and all unknown parameters simultaneously in the PINN-based inverse model, which is similar to the case of identifying the equation parameters of Navier–Stokes equation in [17], where the measured dataset does not include the dependent variable pressure $p$. However, it remains challenging in Giles parameters identification, because internal data other than the input and output are not available as in that case. A more suitable method for the dynamic scenarios is to identify the Giles parameters in steady state and jointly optimized with $N_2(z, t)$ at a lower learning rate under dynamic loading conditions.

To completely model the reaction of an EDFA to transient events, another consideration is the spectral hole burning (SHB) effect, which affects the dynamic response of the EDFA itself and thus potentially influences the reaction of electronic control. The Giles model itself does not take the SHB effect into account, making it difficult for the Giles model-based solutions to simulate the SHB effect. To model SHB effect, changes in spectral power distribution [33] as well as elongated settling time of electronic EDFA control are required to be considered, and a more complex 3-level system incorporating SHB was proposed [34], which introduces more coupled equations and parameters, necessitating further work in terms of measurement and modeling.

In addition, for commercial EDFA devices that package pump light sources and other passive optical structures including isolators, coupler and filters, the gain modeling will not only focus on the EDF itself. In addition to considering all modules together and establishing a mapping between the inputs and outputs of the commercial EDFA in data-driven models, analytical EDFA models requires establishing modules separately and cascading them into a complete system model, which is more accurate and interpretable. Accordingly, the modeling losses of other modules can be added in the total loss, for example in [16], the responses of filters and couplers can be regarded as a constant insertion loss spectrum.

## VII. Conclusion

In this work, we proposed a PIML-based method for physical parameter identification and gain profile estimation of EDFA, which overcame the limitations of both numerical methods and data-driven methods by embedding the physical prior knowledge in the network training process. When physical parameters were known, the PINN-based forward model can be established directly for accurate gain estimation under various input configurations. For most practical scenarios where parameters were unknown or partly known, PINN-based inverse models were first constructed for identifying the physical parameters with only five sets of input-output powers of pump and signal measured at different wavelengths. Then, the PINN-based forward model can be accordingly established for gain estimation using the identified values of parameters. To verify the feasibility of proposed method, the performance in scenarios with known and unknown physical parameters were investigated, respectively. Simulation results showed that PIML can effectively identify the unknown parameters and accurately model the gain profiles of EDFA. Furthermore, an experimental system for signal amplification was constructed to demonstrate the feasibility and performance of the proposed method in practical scenarios. Experimental results demonstrated that the actual values of physical parameters can be effectively identified, which were slightly different from typical values. Compared with using the typical values for gain estimation, smaller MAE of 0.127 dB and STD of 0.065 dB can be achieved using the PINN-based forward model with the identified values. Moreover, we analyzed how the settings of data pairs number and initial values of unknown parameters affect the inverse model. Future works on dynamical loading scenarios and commercial EDFAs are also discussed and prospected. Our work shows that PIML could achieve accurate parameter identification and gain estimation of EDFA with only a few sets of input-output power pairs of pump and signal without intricate measurements or large amounts of data. It is expected to migrate seamlessly to different amplifiers with different doping elements and doping concentrations.